\documentclass[twocolumn, superscriptaddress]{revtex4-1}

\usepackage{lmodern} %Latin Modern = Extended version of Computer modern
\usepackage[T1]{fontenc}
\usepackage[utf8]{inputenc}
\usepackage{microtype}

\usepackage{amssymb,amsmath,braket,cancel}
\usepackage{siunitx}
\usepackage{commath} % includes \dif \abs etc
\DeclareMathOperator{\Tr}{Tr}
 
\newcommand{\ketbra}[2]{\ensuremath{\ket{#1}\!\bra{#2}}}

\usepackage{url}

\usepackage{graphicx}

\begin{document}

\title{Multiplexing Free-Space Channels using Twisted Light}

\author{Brandon Rodenburg}
\email{brandon.rodenburg@gmail.com}
\affiliation{School of Physics and Astronomy, Rochester Institute of Technology, Rochester, New York 14623, USA}

\author{Omar S. Magaña-Loaiza}
\email{omar.maganaloaiza@rochester.edu}
\affiliation{The Institute of Optics, University of Rochester, Rochester, New York 14627, USA}

\author{Mohammad Mirhosseini}
\affiliation{The Institute of Optics, University of Rochester, Rochester, New York 14627, USA}

\author{Payam Taherirostami}
\affiliation{Department of Physics, University at Buffalo, The State University of New York, Buffalo, New York 14260, USA}

\author{Changchen Chen}
\affiliation{The Institute of Optics, University of Rochester, Rochester, New York 14627, USA}

\author{Robert~W.~Boyd}
\affiliation{The Institute of Optics, University of Rochester, Rochester, New York 14627, USA}
\affiliation{Department of Physics, University of Ottawa, Ottawa ON K1N 6N5, Canada}

\date{\today}

\begin{abstract}
    We experimentally demonstrate an interferometric protocol for multiplexing
    optical states of light, with potential to become a standard element in
    free-space communication schemes that utilize light endowed with orbital
    angular momentum (OAM). We demonstrate multiplexing for odd and even OAM
    superpositions generated using different sources. In addition, our
    technique permits one to prepare either coherent superpositions or
    statistical mixtures of OAM states. We employ state tomography to study the
    performance of this protocol, and we demonstrate fidelities greater than
    0.98.
\end{abstract}

\maketitle

\section{Introduction}
Beams of light that are structured in their transverse degree of freedom are
an interesting and powerful tool in quantum information science due the
virtually limitless  level of complexity than can be encoded onto this
structure. One such promising set of modes is the orbital angular momentum
(OAM) states introduced by Allen \emph{et al.}~\cite{Allen1992}. Such modes
have a spiral phase profile $\exp(i\ell\phi)$, where $\phi$ is the azimuthal
transverse angle, and $\ell$ is the unbounded mode index which specifies the
amount of OAM per photon in units of $\hbar$. The usefulness of such modes has
already been demonstrated in communication (both classical and
quantum)~\cite{Gibson2004,Wang2012,Tamburini2012,Mirhosseini2015}, as well as a
fundamental tool in basic quantum information
science~\cite{Mair2001,Molina-Terriza2007,Nagali2009}.

It was recently demonstrated that OAM states, could be transformed to
implement a ``Quantum Hilbert Hotel''~\cite{Potocek2015}. For instance, for
states with OAM index $\ell$, a setup could be built implementing a Hibert
hotel map $\mathbb H_2$ that transforms $\mathbb H_2\ket{\ell} = \ket{2\ell}$.
This means that even if one had a state that contained an infinite amount of
information by utilizing the entire countably infinite OAM basis, more
information could always be added by a transformation of the state, i.e.~an
arbitrary state can be written as $\ket{\psi} = \sum_\ell\psi_\ell\ket{\ell}$,
in which case the Hilbert hotel map would transform this to
\begin{equation}
    \mathbb H_2\ket{\psi} = \sum_\ell\psi_\ell\ket{2\ell},
\end{equation}
leaving the amplitude in all the odd numbered OAM states identically zero.
However, in order to utilize this transformation for quantum information
processing, it is essential to be able to address both the even and odd OAM
subspaces separately and then be able to coherently combine them.

In this paper we experimentally implement such an OAM multiplexer that allows
for the coherent combination of the even and odd order OAM modes as proposed in
Ref.~\cite{Garcia-Escartin2008}. This device interferometrically combines
multiple beams in a manner that is in principle both reversible and completely
lossless, as is necessary for many quantum applications~\cite{Leach2002}. As
has been pointed out, such a multiplexer is also useful for classical
multiplexing of information for use in communication as
well~\cite{Gatto2011,Martelli2011}.

\section{OAM Multiplexer}\label{sec:OAM_Duplexer}
\begin{figure}[htb]
    \centering
    \includegraphics[width=0.49\textwidth]{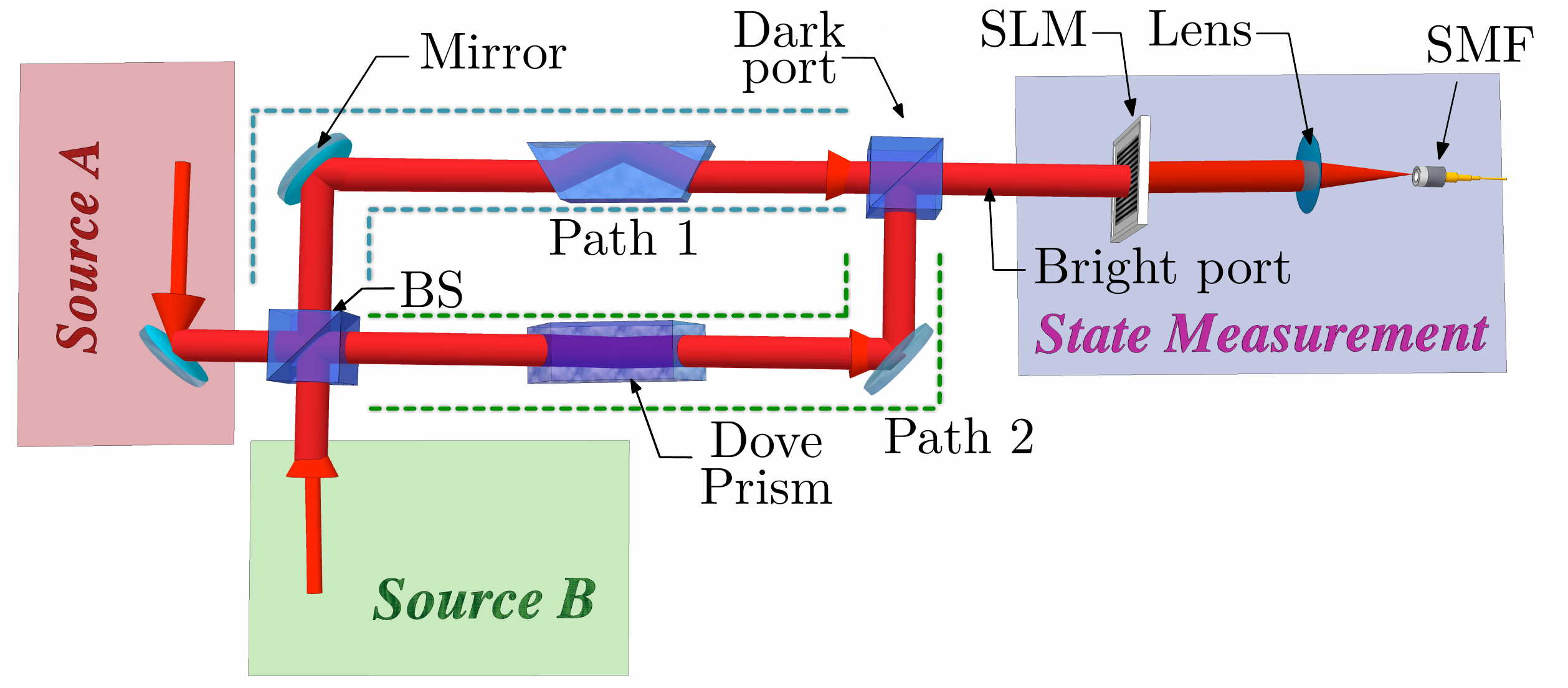}
    \caption{Basic schematic for OAM multiplexer. The device combines the
        symmetric part of input beam $A$ with the antisymmetric part of $B$. }
    \label{fig:OAM_Duplexer}
\end{figure}

\begin{figure*}[htbp!]
    \centering
    \includegraphics[width=0.75\textwidth]{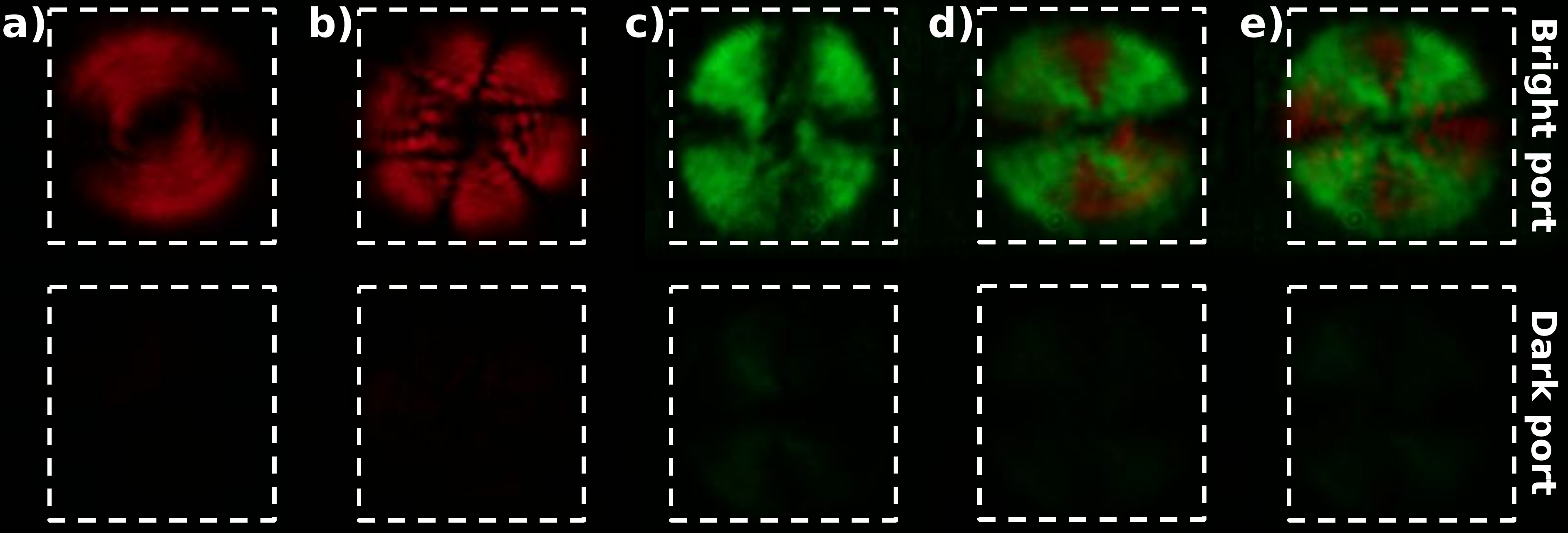}
    \caption{Experimental evidence of the functionality of the OAM duplexer.
        Part a) and b) show the two output ports when only source $A$ is active
        with the antisymmetric states $\ket{\psi_1} = 1/\sqrt{2}(\ket{\ell_1=1}
        + \ket{\ell_2=3})$ and $\ket{\psi'_1} =1/\sqrt{2}(\ket{\ell_1=-1} +
        \ket{\ell_2=5})$ respectfully. Part c) shows the functionality of the
        duplexer when only the second beam is active with symmetric state
        $\ket{\psi_2} = \ket{\ell_3=-2} + \ket{\ell_4=2}$. Finally, parts d)
        and e) show the device's output when the state in c) is combined with
        either a) or c) respectively. We note that the dark port should output
        no light by design, and thus the intensity there is very weak.  We show
        these images to demonstrate that we have very little leakage into this
        dark port, demonstrating the high quality of our setup.} 
\label{fig:output}
\end{figure*}

Consider the interferometric setup in Fig.~\ref{fig:OAM_Duplexer}. The setup consists
of a Mach-Zehnder interferometer with a Dove prism in each arm. Each beam
splitter is a 50:50 beam splitter and the arms of the beam splitter are
arranged such that the output is given as the constructive interference between
paths 1 and 2 for an input at $A$, and destructive interference for an input
from $B$, i.e.
\begin{equation}
    \Ket{f(\mathbf r)}_\text{in}\rightarrow\Ket{f(\mathbf r)}_{out}
        =\frac{1}{\sqrt{2}}\left(\Ket{f_1(\mathbf r)}\pm\Ket{f_2(\mathbf r)}\right),
\end{equation}
where $\Ket{f_{1,2}(\mathbf r)} = \hat{P}_{1,2}\Ket{f(\mathbf r)}_\text{in}/\sqrt{2}$
where $\hat P_{1,2}$ is the net effect of path 1 or 2 on the transverse field
mode $\Ket{f(\mathbf r)}$.

Each reflection causes the spatial mode to experience a parity flip, which for
OAM causes a sign change in the OAM index, i.e.
\begin{equation}
    \Ket{\ell}\rightarrow\Ket{-\ell}.
\end{equation}
Each path has an even number of reflections (including from the prisms) such
that OAM is preserved and the effects of the parity flips effectively cancel.

In addition to a parity flip, the Dove prisms also induce a rotation of the
beam proportional to the angular orientation of the prism itself. The
orientation of the prisms were chosen to be $\pi/2$ relative to each other
creating a relative rotation of $\pi$ between the two beams. This is
represented by letting $\hat P_1 = \hat I$ and $\hat P_2=\hat R_\pi$, where
$\hat R_\pi\Ket{f(r,\phi)} = \Ket{f(r,\phi+\pi)}$.

Now any function $f(\mathbf r)$ can be written as the sum of a symmetric
function $f_S$ and an anti-symmetric function $f_A$ which are eigenstates of
$\hat R_\pi$ with eigenvalues $\pm 1$ respectively. For OAM states, even values
of the OAM index $\ell$ are symmetric states while odd $\ell$ are
anti-symmetric. Now the effect of the setup on a beam from input $A$ becomes
\begin{equation}
\begin{split}
\frac{1}{2}\left(\hat P_1 + \hat P_2\right)&\Ket{f(\mathbf r)}_A\\
    &= \frac{1}{2}\left(\hat I + \hat R_\pi\right)\Ket{f(\mathbf r)}\\
    &= \frac{1}{2}\left(\Ket{f(\mathbf r)}+\Ket{f_S(\mathbf r)}-\Ket{f_A(\mathbf r)}\right)\\
    &= \Ket{f_S(\mathbf r)},
\end{split}
\end{equation}
while the effect for an input from $B$ is
\begin{equation}
\begin{split}
\frac{1}{2}\left(\hat P_1 - \hat P_2\right)&\Ket{f(\mathbf r)}_B\\
    &= \frac{1}{2}\left(\hat I - \hat R_\pi\right)\Ket{f(\mathbf r)}\\
    &= \frac{1}{2}\left(\Ket{f(\mathbf r)}-\Ket{f_S(\mathbf r)}+\Ket{f_A(\mathbf r)}\right)\\
    &= \Ket{f_A(\mathbf r)}.
\end{split}
\end{equation}

Thus the device acts as a filter that outputs the symmetric component of
$\Ket{f(r)}_A$, combined with the antisymmetric component of $\Ket{f(r)}_B$, or
equivalently even and odd OAM states respectively. If $\Ket{f(r)}_A$ is
composed only of even OAM modes, and $\Ket{f(r)}_B$ only contains odd, then
this process is lossless.

\section{Experimental setup and state preperation}\label{sec:Experiment}
Our experimental setup is depicted in Fig.~\ref{fig:OAM_Duplexer}. This scheme
comprises three parts: state preparation, the OAM duplexer and state
measurement. The state preparation consists of two independent sources, a HeNe
laser at \SI{633}{nm} and a solid-state laser at \SI{532}{nm}. Each laser
illuminates a spatial light modulator (SLM) where OAM superpositions are
encoded. 

As a demonstration of our device we prepared two states using our two lasers at
equal intensities, one for each input of the device. Each beam was prepared as
a state within a two dimensional subspace of OAM states (two even and two odd).
The first beam was prepared in a state of the form
\begin{equation}
    \ket{\psi_1} = \alpha_1\ket{\ell_1} + \beta_1\ket{\ell_2},
\end{equation}
and the second laser was prepared in state
\begin{equation}
    \ket{\psi_2} = \alpha_2\ket{\ell_3} + \beta_2\ket{\ell_4},
\end{equation}
where $\ell_{1,2}$ are even and $\ell_{3,4}$ are odd OAM states. Because the two
lasers are incoherent with respect to each other, the expected state at the
output of the device is simply the incoherent sum of the density matrices
formed from $\ket{\psi_1}$ and $\ket{\psi_2}$, i.e.
\begin{equation}
    \hat{\rho} = \frac{1}{2}\left(\ket{\Psi_1}\bra{\Psi_1} + \ket{\Psi_2}\bra{\Psi_2}\right),
\end{equation}
where $\ket{\Psi_1}\propto\ket{\psi_1}\otimes\ket{\psi_2=0}$ and
$\ket{\psi_2=0}$ represents a vacuum state in the space spanned by input
2. Note that
\begin{equation}
    \hat{\rho} \ne \ket{\psi_1}\otimes\ket{\psi_2} 
                   \bra{\psi_1}\otimes\bra{\psi_2},
\end{equation}
which represents a pure state (i.e.~perfect coherence between the two lasers).
Now the density matrix $\rho$ can be represented by a $4 \times 4$ matrix where
the $ij^{th}$ element is represented by
\begin{equation}
    \rho_{ij}\equiv\Braket{\ell_i |\hat{\rho}|\ell_j}
    =\frac{1}{2}\begin{pmatrix}
        |\alpha_1|^2 & \alpha_1\beta_1^* & 0 & 0\\
        \alpha_1^*\beta_1 & |\beta_1|^2 & 0 & 0\\
        0 & 0 & |\alpha_2|^2 & \alpha_2\beta_2^*\\
        0 & 0 & \alpha_2^*\beta_2 & |\beta_2|^2
    \end{pmatrix}.
\label{eqn:Rho}
\end{equation}
Note that $\rho_{ij}=\rho_{ji}=0$ for any combination of $i\in{1,2}$ and
$j\in{3,4}$ due to the incoherence between the lasers and the prepared states
from the two lasers living in separable subspaces of the full Hilbert space.

In order to qualitatively show the functionality of the duplexer we inject
several superpositions. Due to the limited stability of the Mach-Zehnder
interferometer the ratio between the dark and bright port is approximately
only 12\%, although a better dark port can be momentarily obtained if the
alignment is continuously adjusted. The power injected to each port was
approximately \SI{32}{nW}. First we inject the superposition $\ket{\psi_1} =
1/\sqrt{2}(\ket{\ell_1=1} + \ket{\ell_2=3})$, the output is shown in
Fig.~\ref{fig:output}a, later we inject in the same port $\ket{\psi'_1} =
1/\sqrt{2}(\ket{\ell_1=-1} + \ket{\ell_2=5})$ (see Fig.~\ref{fig:output}b). The
even superposition we inject is $\ket{\psi_2} = 1/\sqrt{2}(\ket{\ell_3=-2} +
\ket{\ell_4=2})$, see Fig.~\ref{fig:output}c. In Fig.~\ref{fig:output}d and e,
we demonstrate the action of the duplexer, first we multiplex $\ket{\psi_1}$
and $\ket{\psi_2}$ and later we repeat the experiment with $\ket{\psi'_1}$ and
$\ket{\psi_2}$. As it is shown in Fig.~\ref{fig:output}, most of the light goes
trough port A whereas port B is almost completely dark.

\section{State tomography}\label{sec:measurement}
\begin{figure}[htbp]
    \centering
    \includegraphics[width=0.4\textwidth]{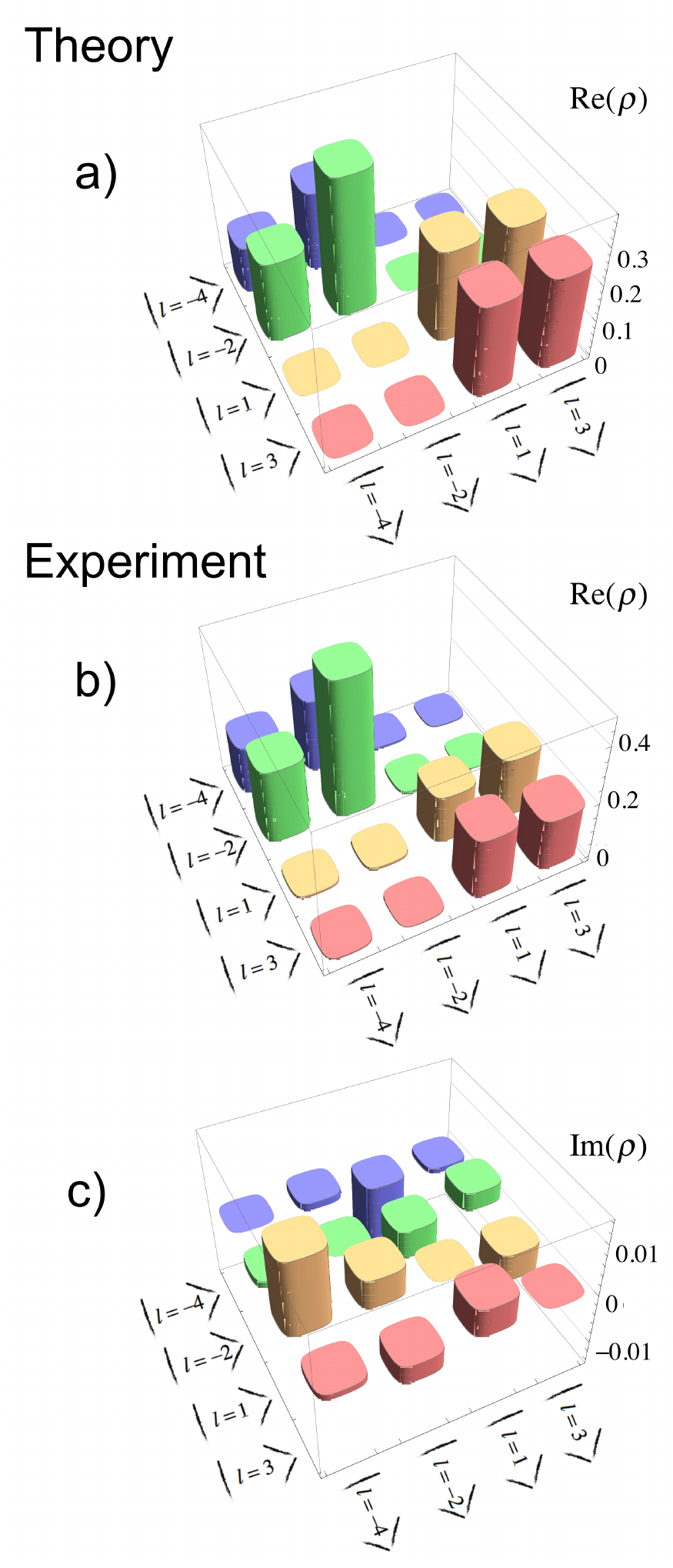}
    \caption{The performance of the duplexer is quantified by means of state
        tomography to measure the output density matrix in order to compare
        this with the intended ideal output given by Eq.~\eqref{eqn:IdealRho}.
        Figure a) shows the ideal density matrix for the injected states, with
        the phases chosen to make the density matrix real. Figure b) shows the
        experimentally reconstructed real part of the density matrix.  Due to
        experimental imperfections, the imaginary part of the density matrix is
        small but non-zero, and is shown in c).}
    \label{fig:results}
\end{figure}

In order to experimentally measure our output $\hat{\rho}$, we need to make different
projection measurements.  If we make a set of projection measurements using a
set of states $\ket{\phi}$, then the measurement
$\hat{\pi}\equiv\ketbra{\phi}{\phi}$ will be found with the following
rate/probability
\begin{equation}
    P=\Tr{(\hat{\pi}\hat{\rho})}=\Braket{\phi|\hat{\rho}|\phi}.
\end{equation}

To measure the subspace spanned by any 2 degrees of freedom (e.g.
$\ket{\ell_1}$ and $\ket{\ell_2}$) we need to make a number of projective
measurements in order to reconstruct the total state $\hat{\rho}$.  Projecting on the
state $\Ket{\phi}=\Ket{\ell_1}$ or $\ket{\ell_2}$ will give the diagonal
elements $\rho_{11}$ and $\rho_{22}$. While
$\ket{\phi}\propto\ket{\ell_1}\pm\ket{\ell_2}$ will give
\begin{equation}
    P_\pm=\Tr{(\hat{\pi}_\pm\hat{\rho})}=\rho_{11}+\rho_{22}\pm(\rho_{12}+\rho_{21}).
\end{equation}

Now $\rho_{12}+\rho_{21}=\rho_{12}+\rho_{12}^*=2\Re{(\rho_{12})}$ so we can get
the real part of $\rho_{12}$ from a differential measurement (to avoid
miscalibration errors)
\begin{equation}
    \Re{(\rho_{12})}=(P_+-P_-)/4.
\end{equation}

In order to find the imaginary part $\Im{(\rho_{12})}$, we need to measure in a
3rd basis.  This is why in quantum tomography of a qubit 
one needs to measure in $\hat{\sigma}_x$, $\hat{\sigma}_y$, and $\hat{\sigma}_z$ bases, or
equivalently measure the three Stokes Parameters $S_1$, $S_2$, and $S_3$. For
this reason our measured parameters $P$ are sometimes referred to as ``Qudit
Stokes Parameters''~\cite{Altepeter2005}.

So to get the imaginary part of $\rho_{12}$, we measure
$\ket{\phi}=\ket{\ell_1}\pm i\ket{\ell_2}$ and follow a similar procedure as
before which gives
\begin{equation}
    P'_\pm =
    \Tr{(\hat{\pi}_\pm\hat\rho)}=\rho_{11}+\rho_{22}\pm i(\rho_{12}-\rho_{21}).
\end{equation}
Taking the difference of these two rates allows one to find the imaginary part
of $\rho_{21}$ which is given by
\begin{equation}
    \Im{(\rho_{12})}=(P'_--P'_+)/4.
\end{equation}

In order to make our projection measurements, we use the standard method
developed for measuring spatial modes~\cite{Mair2001}. The output of the bright
port of our device was imaged onto an SLM with the complex conjugate of the
field mode we wish to measure, encoded onto a modulated diffraction grating.
We then couple the first diffraction order into single mode fiber and measure
the count rate on an APD. It has been demonstrated that for such projection
measurements there exists cross-talk between neighboring
modes~\cite{Qassim2014}. In order to avoid this issue we prepared the
incoherent superposition of $\ket{\psi_1} = 1/\sqrt{4}\ket{1} +
\sqrt{3/4}\ket{3}$ and $\ket{\psi_2} = 1/\sqrt{2}(\ket{-2} + \ket{-4})$, which
allows us to obtain cleaner results in our projection measurment as neighboring
modes are not used. Written in matrix formation for the basis states
$\ell\in\left[-4, -2, 1, 3\right]$ gives [via Eq.~\eqref{eqn:Rho}]
\begin{equation}
\rho_{ij} = \frac{1}{8}
\begin{pmatrix}
2  &  2  &      0     &      0  \\
2  &  2  &      0     &      0  \\
0  &  0  &      1     &  \sqrt{3}  \\
0  &  0  &  \sqrt{3}  &      3  \\
\end{pmatrix}.
\label{eqn:IdealRho}
\end{equation}

Our results are shown in Fig.~\ref{fig:results}. The phases of our states where
chosen such that $\rho$ is ideally real as shown in Fig.~\ref{fig:results}a.
Our measured state, shown in Fig.~\ref{fig:results}b (c) shows the real
(imaginary) part of our measure state $\rho_m$, which demonstrates excellent
agreement with our intended state. Using the standard measure of fidelity
defined as~\cite{Jozsa1994}
\begin{equation}
    F(\rho,\rho_m)\equiv\Tr\left[\sqrt{\rho}\rho_m\sqrt{\rho}\right]^{1/2},
\end{equation}
we find our measured state has a fidelity of 0.9880. 

In conclusion, we have experimentally demonstrated the use of a protocol for
multiplexing even and odd spatial modes of light with OAM states. We have shown
that this multiplexing scheme can generate general incoherent mixes of states
in a simple and deterministic way. The fact that this protocol works at low or
single photon levels makes this scheme a promising tool for use in quantum
information tasks.

\section{Acknowledgements}
The authors thank E.~Karimi for excellent discussion. RWB acknowledges funding
from the Canada Excellence Research Chairs program. OSML acknowledges support
from the Consejo Nacional de Ciencia y Tecnología (CONACyT), the Secretaría de
Educación Pública (SEP), and the Gobierno de Mexico.

\end{document}